\documentclass[prl,twocolumn,floatfix,showpacs,amsmath]{revtex4}
\usepackage{amssymb}
\usepackage{graphicx}

\begin{document}
\newcommand{\be}{\begin{eqnarray}}
\newcommand{\ee}{\end{eqnarray}}
\newcommand{\bea}{\begin{eqnarray}}
\newcommand{\eea}{\end{eqnarray}}
\newcommand{\bma}{\begin{subequations}}
\newcommand{\ema}{\end{subequations}}
\def\lR{l^2_{\mathbb{R}}}
\def\N{{\cal N}}
\def\H{{\cal H}}
\def\B{{\cal B}}
\def\P{{\cal P}}
\def\dd{\delta}
\def\tr{\textmd{tr}}
\def\one{{\bf 1}}

\title{Renormalization algorithm for the calculation of spectra of interacting quantum systems.}

\author{D. Porras$^1$}

\author{F. Verstraete$^{1,2}$}

\author{J. I. Cirac$^{1}$}
\affiliation{
$^1$ Max-Planck Institut f\"ur Quantenoptik,
Hans-Kopfermann-Str. 1, Garching, D-85748, Germany. \\
$^2$ Institute for Quantum Information, Caltech, Pasadena 91125, CA.}

\date{\today}

\begin{abstract}

We present an algorithm for the calculation of eigenstates with definite linear momentum in quantum lattices. Our method is related to the Density Matrix Renormalization Group, and makes use of the distribution of multipartite entanglement to build variational wave--functions with translational symmetry. Its virtues are shown in the study of bilinear--biquadratic $S=1$ chains.

\end{abstract}
\maketitle
%
%importance of numerical methods and DMRG, conexion to quantum information, new methods, some %problems of DMRG
Quantum many--body systems pose problems of great complexity that only in a few cases can be solved analytically. For this reason numerical algorithms have played a decisive role in understanding strongly--correlated matter. 
A remarkable example is the Density Matrix Renormalization Group (DMRG), which is a powerful tool for the description of ground state properties \cite{White}. 
On the other hand, the merging of Quantum Information Theory \cite{NielsenChuang} and Condensed Matter has given us a new insight into the physics of interacting quantum systems. The theory of entanglement has yielded new tools to quantify quantum correlation \cite{Vidal.Kitaev}, as well as a new theoretical framework for DMRG \cite{PBC}, and helped us to develop algorithms to deal with problems in higher dimensions \cite{PEPS}, and also to describe time evolution, systems at finite temperature and quantum dissipation \cite{DMPS,Vidal04}.

DMRG is a variational method over the class of Matrix Product States \cite{RomerPRL,PBC}, which correspond to the 1D realization of the more general Projected Entangled--Pair States (PEPS) introduced in \cite{PEPS}. 
In a PEPS each site in a lattice is described by a set of auxiliary systems which form entangled pairs with their first neighbors, and the physical state is built by local maps onto the physical Hilbert space. 
The distribution of bipartite entanglement governs the characteristics of PEPS, which are ideally suited to describe systems with short--range correlations, something that explains why DMRG is particularly accurate in describing non--critical ground states. 
This observation invites us to consider the intriguing possibility of modifying the auxiliary state underlying PEPS to distribute multipartite entanglement and build new variational classes that are more suitable for a given problem.

In this letter we present the following results: 
(i) We define the Projected Entangled--Multipartite States (PEMS) and study the particular case in which a GHZ--like state is added to the auxiliary system underlying PEPS. The resulting variational states have a given definite linear momentum $k$. (ii) The new variational class is used to describe efficiently excitations of translational invariant Hamiltonians by means of a numerical DMRG--like algorithm. In this way, we can calculate the lowest energy branch of excitations, that is, the set of minimum energy eigenstates for different linear momenta $\{|\Psi^{[0]}_k\rangle\}$. (iii) We also present an algorithm for the calculation of the sequence of excited states at a given point in momentum space, $\{ | \Psi_k^{[0]} \rangle, | \Psi_k^{[1]} \rangle, \ldots \}$, which has a broad usefulness, and can also be implemented together with non-translational invariant DMRG--like algorithms. (iv) The utility of the method is shown in the study of bilinear--biquadratic S=1 spin chains, where we find indications of a quantum phase characterized by nematic quasi--long range order, which can be realized in experiments with cold atoms in optical lattices \cite{S1.optical.lattice,Imambekov}. 

We introduce our method by considering the case of a chain of $N$ $d_s$--dimensional spins. Let us assign a set of auxiliary subsystems $x_n$ to each site $n$. In 1D PEPS, only auxiliary systems corresponding to adjacent sites, say $x_n$ and $x_{n+1}$, are entangled in $|\Psi_{aux}\rangle$. To describe efficiently multipartite entanglement, such as the one present in critical or spin--wave--like states, one should consider the most general case in which $|\Psi_{aux} \rangle$ is multipartite entangled, that is, it cannot be reduced to a product state of pairs of entangled sites \cite{note0}. The physical wave--function is created with the aid of a product of local maps $P_n$:
\be
|\Psi_{PEMS} \rangle = P_1 \otimes P_2 \otimes \ldots \otimes P_{N} \ 
| \Psi_{aux} \rangle
\ee 
Each $P_n$ is a local map from $x_n$ to the spin Hilbert space at site $n$.
Let us see how the choice of the proper $| \Psi_{aux} \rangle$ allows us to find a variational class with translational invariance. Each site $n$ is described with the aid of two auxiliary subsystems, $a_n$, $b_n$ of dimension $D$ and a third subsystem $c_n$ of dimension $N$. $a_n$, $b_{n-1}$ form a maximally entangled state, $| \phi \rangle$. On the other hand,
the $c_n$ are in the following multipartite entangled state \cite{note1}:
\be
| M_k \rangle \equiv \frac{1}{\sqrt{N}} \sum_{n=0}^{N-1} 
e^{i k n} T_n |1\rangle_{c_1} |2\rangle_{c_2} \ldots |N\rangle_{c_N} ,
\label{definition.SW}
\ee
where $k = n_k 2 \pi / N$, with $n_k = 0, \ldots, N-1$, and $T_n$ is the translation operator. The auxiliary state is, thus, $| \Psi_{aux} \rangle = |\Phi\rangle^{\otimes N}|M_k\rangle$ (see Fig. \ref{fig.scheme}). Note that, due to the multipartite entanglement in $| \Psi_{aux} \rangle$, subsystems $c_n$ are long--range correlated, such that, after the local mapping, the corresponding variational wave--function is better suited to describe long--range correlation. 

In this case, contrary to the non-translational ansatz in \cite{PBC}, all $P_n$'s are given by the same operator acting on different sites of the chain, so that their product is a translational invariant operator. The local maps are determined by a set of $N d_s$ matrices of dimension $D$, 
$\{ A^s_{[\gamma]} \}_{\substack{s=1,\ldots,d_s \\ \gamma = 1,\ldots,N}}$:

\begin{figure}[t]
  \center
  \includegraphics[width=\linewidth]{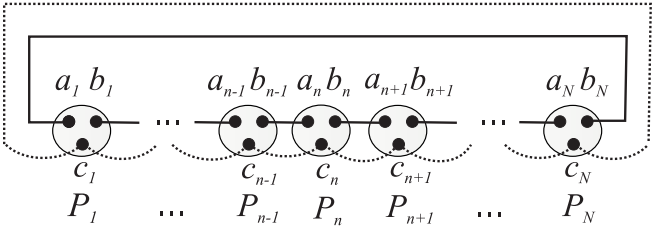}
  \caption{Schematic picture of the Projected Entangled--Multipartite States presented in the text.}
  \label{fig.scheme}
\end{figure}

\be
P = \sum_{s,\alpha,\beta,\gamma} (A^s_{[\gamma]})_{\alpha,\beta}
\ | s \rangle_{\ a \!}
\langle \alpha |_{\ b \!} \langle \beta |_{\ c \!} \langle \gamma | ,
\label{definition.projector}
\ee
such that $P_n$ acts on the auxiliary states of site $n$ and returns the spin state $| s \rangle$ with amplitude $(A^s_{[\gamma]})_{\alpha,\beta}$, provided $a_n$, $b_n$, $c_n$ are in states $|\alpha \rangle$, $|\beta \rangle$, $|\gamma \rangle$, respectively.   
The resulting physical states form a variational class of translational invariant states with momentum $k$:
\be
| \Psi_k \rangle = \hspace{-0.25cm}
\sum_{n=0, \{s_j\}}^{N-1} \hspace{-0.1cm}
\frac{e^{i k n}}{\sqrt{N}} T_n \textmd{tr} \{A^{s_1}_{[1]} A^{s_2}_{[2]} \ldots A^{s_N}_{[N]} \}  |s_1\rangle \ldots |s_N\rangle .
\label{MPS1}
\ee
The physical meaning of $\gamma$ becomes clear in Eq. (\ref{MPS1}). 
Due to the presence of  $|M_k \rangle$ the auxiliary state is mapped onto a linear combination of 1D PEPS formed by circular permutations of $A^{s}_{[\gamma]}$, in which $\gamma$ represents the site in the chain. 
%Note that (\ref{MPS1}) can be defined on lattices of any spatial dimension $D_s$, by considering first the $D_s$ structure of PEPS presented in \cite{PEPS}, and adding $N^{D_s}$--dimensional auxiliary subsystems $c_{\vec{n}}$, at each site $\vec{n}$. The auxiliary state $|M_{\vec{k}}\rangle$ is defined then by replacing $k$, $n$ by $D_s$--vectors in (\ref{definition.SW}).  

Averages of observables are computed efficiently with the aid of the following matrices of dimension $D^2$:
\be
E_{O}^{n,d} = \sum_{s,s'}
\left( A^{s}_{[n]} \otimes (A^{s'}_{[n-d]})^{*} \right) \langle s' | O | s \rangle .
\label{positive.maps}
\ee
From now on 
$\pmod{N}$ 
%${\textmd{mod}} \ N$ 
is implicit in all functions of $n$, $d$. The expectation value of any operator is a linear combination of $N$ products of $D^2 \times D^2$ matrices:
\be
\langle O_1 O_2 \ldots O_N \rangle = \frac{1}{N} \sum_{n,d} e^{-i k d}
\textmd{tr} \{ E_{O_1}^{n,d} E_{O_2}^{n+1,d} \ldots E^{n-1,d}_{O_N} \}.   
\label{mean.expectation}
\ee
Eq. (\ref{mean.expectation}) shows that the calculation of averages is decomposed into a sum over $N$ Fourier components.

Let us see how to find the state of the form (\ref{MPS1}) that minimizes the energy of a given Hamiltonian. 
We consider for concreteness the case of a short range spin model, 
$H = \sum_{\mu,n} g_\mu \sigma^\mu_n \sigma^\mu_{n+1}$, where $\sigma^\mu$ form an orthogonal set of hermitian operators. 
The norm and the mean value of the energy can be calculated with the aid of Eq. (\ref{mean.expectation}):
\be
\langle \Psi_k | \Psi_k \rangle &=& 
\sum_{d} e^{-i k d} \tr \{E^{1,d}_\one \dots E^{N,d}_\one\}
 \\
\langle \Psi_k | H | \Psi_k \rangle &=& 
\sum_{n,d,\mu} g_\mu e^{-i k d} 
\tr \{ E^{1,d} \hspace{-0.2cm} \dots E^{n,d}_{\sigma_\mu} E^{n+1,d}_{\sigma_\mu} \hspace{-0.2cm} \dots E^{N,d}_\one \} \nonumber
%\nonumber \\
%\left. + E^{1,d}_{\sigma_\mu} E^{2,d}_\one \dots E^{N-1,d}_\one  E^{N,d}_{\sigma_\mu} \right) .
\label{bilinear.norm}
\ee   
Any expectation value calculated with (\ref{MPS1}) is a bilinear form of each $A^s_{[n]}$ separately, $\langle \Psi_k | \Psi_k \rangle = A^{\dagger}_{[n]} \N[n] A_{[n]}$, 
$\langle \Psi_k | H | \Psi_k \rangle = A^{\dagger}_{[n]} \H[n] A_{[n]}$, where $A_{[n]}$ is the vector obtained by contracting $s$, $\alpha$, and $\beta$ in a single index.
Thus, the energy can be minimized with respect to the set of matrices $A^s_{[n]}$ with a given $n$, by solving the generalized eigenvalue problem $\H[n] A_{[n]} = \epsilon_0 \N[n] A_{[n]}$. This fact allows us to find the optimum PEMS in an iterative way that is similar to other DMRG--like methods: once we have found the optimum $A^s_{[n]}$, we replace it in the wave-function (\ref{MPS1}), then we actualize $\H[n+1]$, $\N[n+1]$, and repeat the minimization with respect to $A^s_{[n+1]}$. The process is repeated in several sweeps from $n=1$ to $N$, until the energy converges.

According to Eq. (\ref{mean.expectation}) $\H[n]$ and $\N[n]$ in (\ref{bilinear.norm}) can be expressed as ${\N}[n] = \sum_d e^{-i k d} {\N}[n,d]$, ${\H}[n] = \sum_d e^{- i k d} {\H}[n,d]$,
where $\N[n,d]$, $\H[n,d]$ depend on matrices $E_O^{m,d}$. In order to find an explicit expression for the bilinear forms, we notice first that matrices $A^s_{[n]}$ appear twice in each product in Eq.(\ref{mean.expectation}), both in $E^{n,d}_O$ and $E^{n+d,d}_{O}$, for example:  
\be
\N[n,d]_{
(^{\hspace{0.1cm} s}_{\alpha \beta}\!),
(^{\hspace{0.1cm} s'}_{\alpha' \beta'}\!)} \hspace{-0.5cm} &&= \tr \{ \!
(  \chi[^{\alpha'}_{\beta'}] \! \otimes \! (\!A_{[n+d]}^{s'}\!)^*  ) E^{n+1,d}_\one \hspace{-0.3cm} \dots E^{n+d-1,d}_\one \times \nonumber \\
&& \hspace{-1.cm}
( A^{s}_{[n-d]} \! \otimes \! \chi[^{\alpha}_{\beta}] ) E^{n+d+1,d}_\one \! \! \! \! \ldots E^{n-1,d}_\one \},
\ee
where $\chi[^{\alpha}_{\beta}]$ is a $D \times D$ matrix with all elements $0$, but a $1$ in the entry $(\alpha, \beta)$.
A similar expression can be found for $\H[n,d]$, which includes a sum over $N$ terms that corresponds to each interaction in the Hamiltonian, and involves also products of $E^{n,d}_{\sigma_\mu}$. 

In principle $\H[n,d]$ and $\N[n,d]$ could be calculated at each step by means of (\ref{positive.maps}, \ref{mean.expectation}). However the structure of the problem resembles that of the non--translational invariant calculation and is suitable for a procedure of initialization and actualization of block operators similar to the one presented in \cite{PBC}. First, we note that there are $N$ types of products corresponding to each Fourier component in (\ref{mean.expectation}). 
$\H[n,d]$ requires $N$ products (from the $N$ interacting terms in the Hamiltonian) of $N$ matrices. A naive estimation would thus yield $N^3$ multiplications at each step during the optimization, but we can speed up the algorithm by storing products of matrices $E^{n,d}_O$ in the proper way. 
For example, for building the matrix $\N[1,d]$, we need the product $E^{2,d}_\one E^{3,d}_\one \ldots E^{d,d}_\one$, which is obtained by calculating, and storing, a sequence of products that we label left operators: 
$E^{d,d}_\one$, $E^{d-1,d}_\one E^{d,d}_\one$, \dots . 
In the next step, we need
$E^{3,d}_\one \ldots E_\one^{d+1,d}$, and we compute it by multiplying the stored left operator $E^{3,d}_\one \ldots E^{d,d}_\one$, and $E^{d+1,d}$, which is the first of a set of products that we label right operators.
In the following, to build $\N[n,d]$, we use the stored left operators, as well as a new right operator $E^{d+1,d}_\one \ldots E^{d+n-1,d}_\one$, that is obtained by a single matrix multiplication from the right operator of the previous step. 
Finally, at $n=d$, we have to calculate and store again all the left operators, and the procedure starts again. A similar recipe can be used for the calculation of $\H[n,d]$, which relies on the use of recursive relations between operators involving matrices $E^{n,d}_{\sigma_\mu}$ \cite{PBC}, and the time for one optimization step is reduced to scale like $N$ (see the Appendix for a more detailed explanation).  
Thus, an increase of the order of $N$ in computation time is required by this algorithm in comparison with the non--translational invariant case presented in \cite{PBC}.
On the other hand if we write (\ref{MPS1}) as a Matrix Product State, we need $D' = D N$ dimensional bonds, something that indicates that, in order to get the same accuracy than in the non--translational schemes, our method requires a lower $D$.

In the following we show how to modify our method to obtain the sequence of excited states with a given linear momentum. Let us consider that we have previously calculated the $M$ lowest energy eigenstates, $|\Psi^{[j]}_k\rangle$, with linear momentum $k$, described by the set of matrices ${B[j]^s_{[\gamma]}}$. In order to proceed further and find the $M+1$ lowest energy state, we would like to run the optimization algorithm under the following constraint:
\be
\langle \Psi_k^{[j]} | \Psi_k \rangle = 0, \ \ \ \ \forall \ j = 1, \ldots, M.
\label{constraint}
\ee
Within the variational class of 1D PEMS defined by (\ref{MPS1}), each constraint in Eq. (\ref{constraint}) is given by:
\be
\langle \Psi_k^{[j]} | \Psi_k \rangle =  \sum_{d=0}^{N-1} 
e^{-i k d}\tr \{ E^{1,d}_{[j]} E^{2,d}_{[j]} \ldots E^{N,d}_{[j]}\} , 
\label{constraint.MPS}
\ee  
where we use the definition:
\be
E^{n,d}_{[j]} = \sum_s A^s_{[n]} \otimes (B[j]^s_{[n-d]})^* .
\label{positive.maps.constraint}
\ee
Eq. (\ref{constraint}) implies a linear constraint in the optimization with respect to each $A^s_{[n]}$ separately.
% which is equivalent to the restriction of the optimization to the subset of PEMS that are orthogonal to the set $\{|\psi_k^{[j]} \rangle \}_{j=1,\ldots,M}$. 
In order to include this constraint in our algorithm we proceed as follows. At each step $n$ we use (\ref{positive.maps.constraint}) to calculate the linear form $\B[j,n]$ that imposes the orthogonality to $|\Psi^{[j]}_k \rangle$ in terms of $A^s_{[n]}$:
\be
\langle \Psi^{[j]}_k | \Psi_k \rangle =  \sum_{s,\alpha,\beta}
({({\B}[j,n])^{s}_{\alpha,\beta}})^* (A^{s}_{[n]})_{\alpha,\beta} = 0 .
\label{linear.constraint}
\ee
${\B}[j,n]$ can also be decomposed into Fourier components, in the same way as $\H[n]$, $\N[n]$ , and each of these components can be computed by the same procedure for actualization and storage of block operators that was explained below. If we contract ($s$, $\alpha$, $\beta$) in a single index, then (\ref{linear.constraint}) reads $B[j,n]^{\dagger} A_{[n]} = 0$. 
The linear constraint is incorporated to the optimization procedure by defining projectors in the subset of states that are orthogonal to the $M$ lowest states, that is, 
$\P[n] = \one - \sum_{i,j} B[i,n] (N_B^{-1})_{i,j} B[j,n]^{\dagger}$, with $(N_B)_{i,j}$ = $B[i,n]^{\dagger}B[j,n]$,  and solving the eigenvalue problem defined by the new Hamiltonian and norm matrices given by $\H[n] \rightarrow \P[n] \H[n] \P[n]$, $\N[n] \rightarrow \P[n] \N[n] \P[n]$. The same idea can be used in other DMRG--related algorithms.
% because orthogonality to a set of lower energy states can always be expressed as a linear constraint at each step in the optimization.

We have applied our method to the study of
bilinear--biquadratic $S=1$ chains, 
$H = \sum_i h_{i,i+1} = \sum_i \cos \theta \vec{S}_i \vec{S}_{i+1} + 
\sin \theta (\vec{S}_i \vec{S}_{i+1})^2$, which display a rich variety of phases depending on the parameter $\theta$.
We focus here on the region $-3\pi/4 \leq \theta \leq -\pi/2$, whose characteristics have been yet not fully understood.
The two limits, $-\pi/2$, $-3 \pi/4$, are exactly solvable and correspond to a gapped dimerized \cite{DimerPhase}, and a gapless ferromagnetic phase \cite{Ortiz}, respectively. 
So far, it has remained unclear what happens between the dimerized and ferromagnetic phases, in particular, whether the dimer order survives down to $\theta = -3 \pi/4$. In Ref. \cite{Chubukov}, it was conjectured that a 1D quantum nematic phase should appear as an intermediate phase. Since then, a few numerical works have dealt with this problem \cite{Fath,recent.DMRG,Kawashima}, but this phase is yet not fully characterized.  

The ability to calculate excitations in a controlled way, makes our algorithm ideally suited to deal with this problem.
Fig. \ref{fig2} shows the spectrum of low energy states at two different points in the phase diagram. 
At $\theta = - \pi/2$ the dispersion relation corresponds qualitatively to a gapped phase. Note that for the calculation of the spectrum it is necessary to include the constraint (\ref{constraint}), in order to find the second lowest energy state at points $k=0$, $k = \pi$. At $\theta = - 0.74 \pi$, which lies within the conjectured quantum nematic phase, the spectrum shows a qualitative change that involves the appearance of a soft mode. The convergence of the algorithm with $D$ is shown in Fig. \ref{fig2} (a). At $\theta = -\pi/2$ we compare our results for the excited state $E^{[1]}_{k=0}$ with exact results obtained by Bethe-ansatz \cite{DimerPhase}. The absolute error in the ground state, calculated by the extrapolation of $E^{[0]}_0$ to $D=12$ ($\Delta E^{[0]}_0 = 2 \times 10^{-3}$) agrees with the error in the first excited state ($\Delta E^{[1]}_0 =  3 \times 10^{-3}$) obtained from the comparison with the exact result at $\theta = -\pi/2$. 

\begin{figure}
  \center
  \includegraphics[width=1.6in]{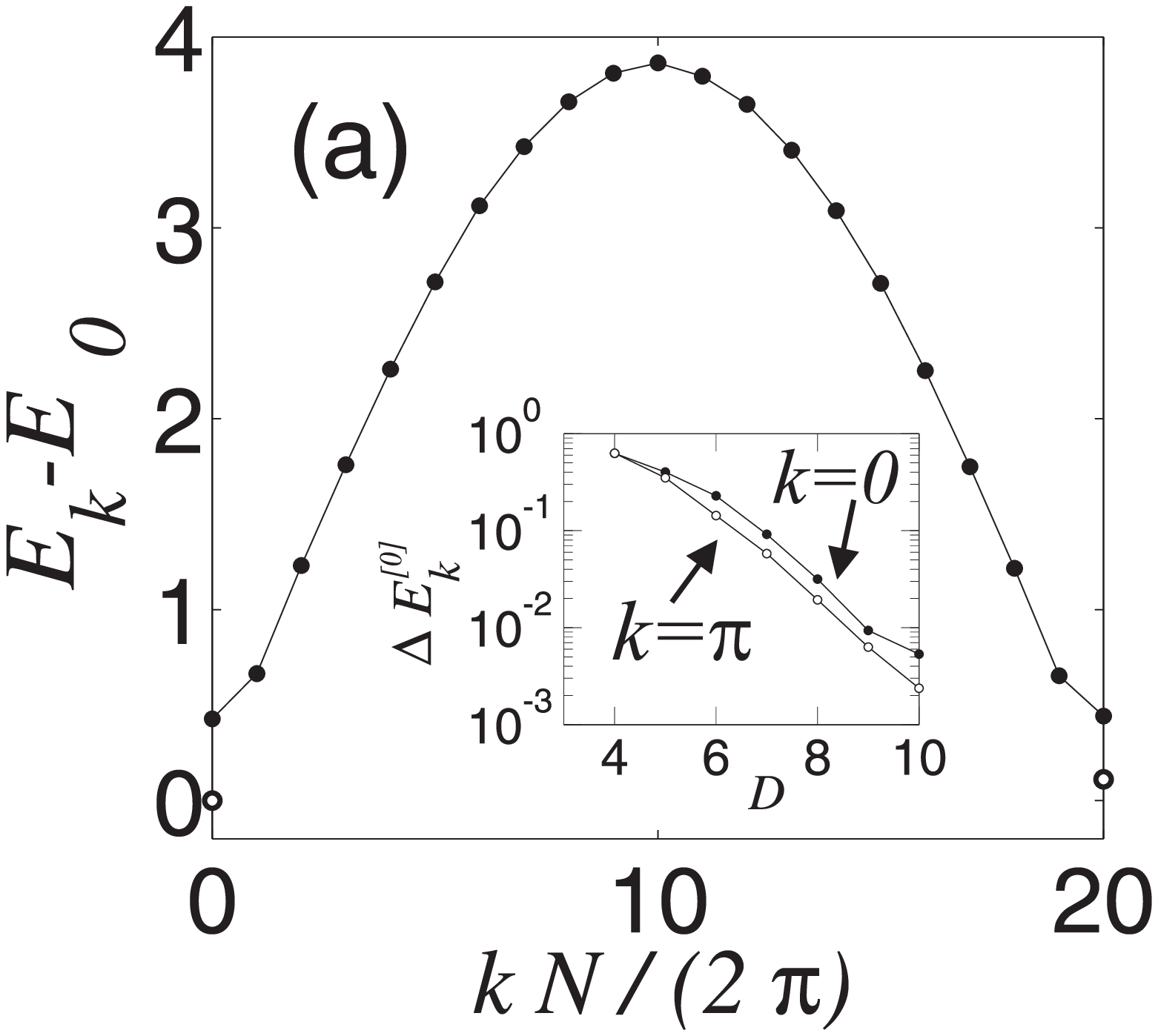}
  \hspace{0.1cm}
  \includegraphics[width=1.5in]{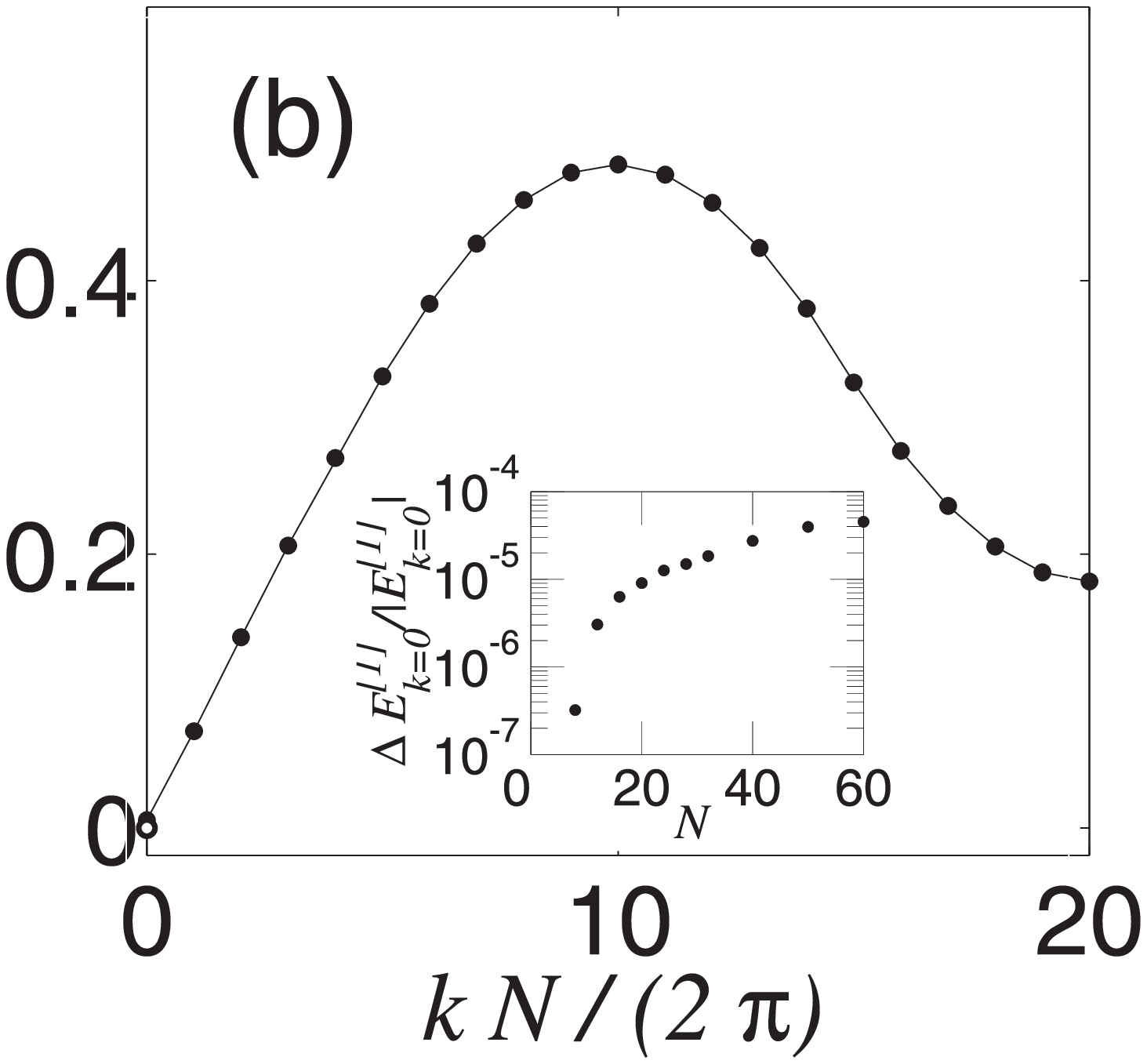}
  \caption{Lowest states of a bilinear--biquadratic $S=1$ chain, $N=40$ sites, $D=10$. 
(a) $\theta = -\pi/2$, $E_0 = -2.7976 N$, (b) $\theta = - 0.74 \pi$, $E_0 = -1.4673 N$. 
Empty circles: lowest energy states. Filled circles: first branch of excitations. Estimated absolute error $\Delta E_k \approx \ 5 \times 10^{-3}$. Inset (a): error in the absolute energies $E^{[0]}_{0}$, $E^{[0]}_{\pi}$ as a function of $D$, estimated by comparison with $D=14$ calculations, for $\theta = -0.5 \pi$. Inset (b): relative error as a function of $N$ in the first excited state energy with $k=0$, $\theta = -\pi/2$.}
\label{fig2}
\end{figure}

By means of Eq.(\ref{mean.expectation}) we can also calculate order parameters (OP) in the ground state. Long--range dimer order is characterized by a non--zero value of
$\langle {\cal D}^2 \rangle/N^2$ in the thermodynamical limit, where  
${\cal D} = \sum_{i}( -1)^i h_{i,i+1}$ is the bond--strength oscillation. 
In the interval $\theta \leq \theta_c$, $\theta_c \approx - 0.7 \pi$, our finite--size results are extrapolated to a value
${\cal D}^2/N^2 < 3 \times 10^{-5}$, which is set by our accuracy,  estimated by comparison with higher $D$ calculations  (Fig. \ref{fig3} (a)). On the other hand the nematic OP is given by the quadrupole tensor, whose components are rotations of ${\cal Q}^{zz} = \sum_i \left( (S_i^z)^2 - 2/3 \right)$.  
Long--range nematic order is described by the isotropic squared OP, 
${\cal Q}^2 = \int d\Omega \langle \left({\cal Q}^{zz}_\Omega \right)^2 \rangle$, 
where
${\cal Q}^{zz}_\Omega$ is ${\cal Q}^{zz}$ rotated to the solid angle $\Omega$. We find that ${\cal Q}^2(N) \propto N^{\alpha_{nem}}$ in the interval $-3\pi/4 < \theta < \theta_c$ with  $1.4 < \alpha_{nem} < 2$ (see Fig. \ref{fig3} (b)). Thus, long--range nematic order is absent, in qualitative agreement with Coleman's theorem \cite{Coleman}.
The fact that ${\cal Q}^2$ decays algebraically with $\alpha_{nem} > 1$ is consistent with the existence of quasi--long range order, as defined by algebraic decay of nematic correlation functions \cite{Auerbach}. Note that $\alpha_{nem}$ evolves continuously to the value $\alpha_{nem} = 2$ in agreement with the exact solution at $\theta = -3\pi/4$ \cite{Ortiz}.

The spectrum in the thermodynamical limit can be accurately determined by studying the scaling of the gaps with system size. 
In particular, the gap between $k=0$, and $k=\pi$ is extrapolated to zero within our numerical accuracy ($\Delta E < 10^{-3}$) in the whole region under study (Fig. \ref{fig.4} (a)). 
%Note that the ability to target the $k = \pi$ state is extremely useful here because one does not need to calculate all the %intermediate low energy states, something that hindered the application of DMRG to this problem \cite{Fath}.
We have also studied the scaled gap between the lowest $k=0$ states, $N \left(E^{[1]}_0 - E^{[0]}_0 \right)$. This quantity should grow linearly with $N$ in a gapped phase, however for values $\theta \leq \theta_c$ the $k=0$ the scaled gap saturates (Fig. \ref{fig.4} (b)). Our results in the range $\theta \leq \theta_c$, $\theta_c \approx - 0.7 \pi$ are thus consistent either with (i) a gapless quantum phase with nematic quasi--long range order, (ii) a phase with correlation lengths longer than the size of the chains considered here, in which case Fig. \ref{fig.4} would not correspond to the asymptotic regime, or (iii) a gap that is smaller than our numerical accuracy.

In conclusion, we have presented a DMRG-like variational algorithm that allows us to find the lowest energy states
with a definite momentum of translational invariant Hamiltonians. 
The variational class of states we used in the algorithms was obtained by extending the concept of Matrix Product
States / Projected Entangled Pair States to include a particular multipartite entangled state. An interesting extension of this work is to explore how other multipartite states with long--range correlations could help in simulating e.g. critical systems.
 
\acknowledgements We thank M.A. Mart\'{\i}n Delgado, G. Ortiz, E. Demler, E. Altmann and S. Montangero for interesting discussions. Work supported by DFG (SFB 631), European
projects, Bayerischen Staatsregierung Quanteninformation, MEIF-CT-2004-010350 and the
Gordon and Betty Moore Foundation (the Information Science and Technology Initiative, Caltech).

\begin{figure}[t]
  \center
  \includegraphics[width=1.5in]{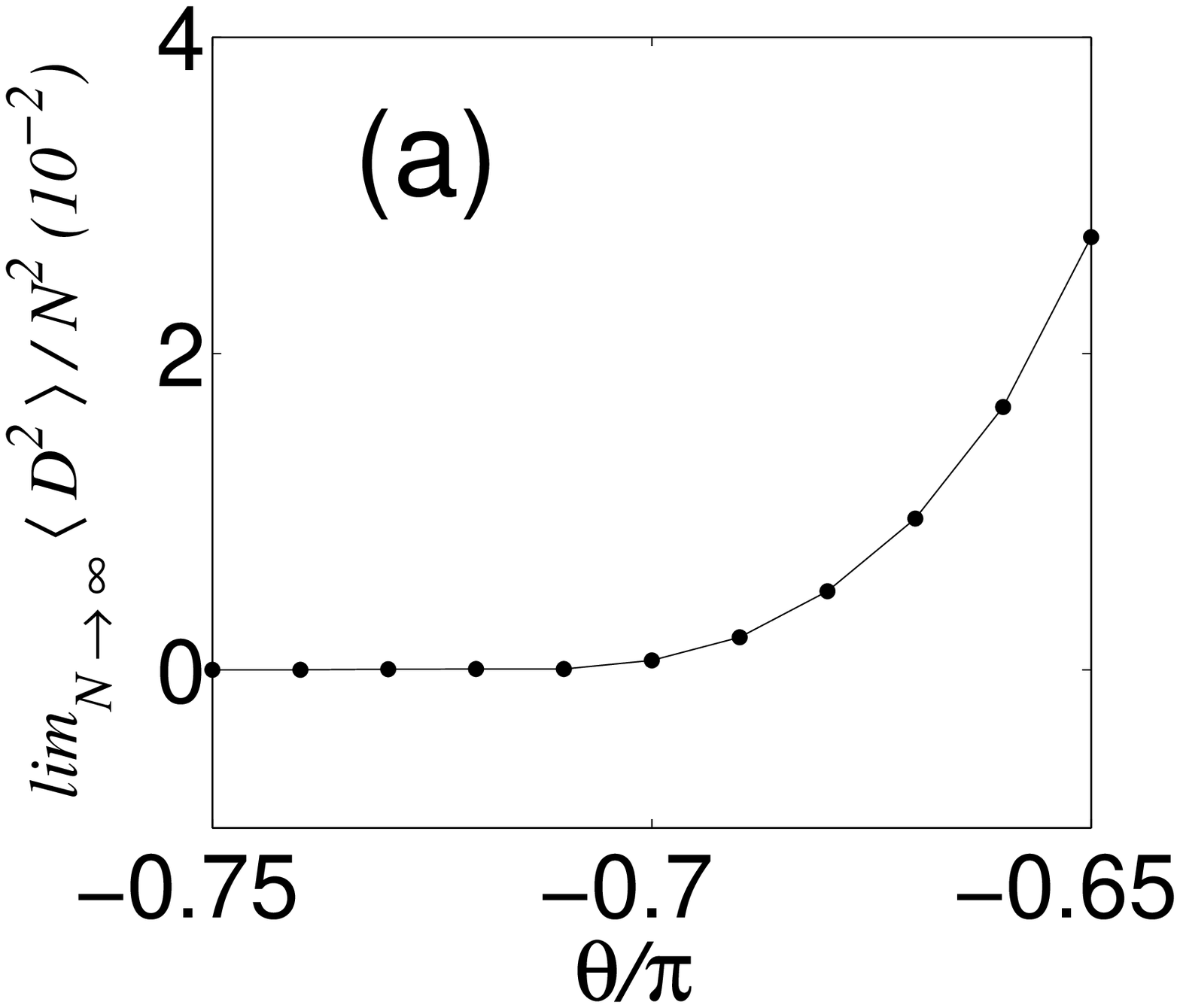}
  \includegraphics[width=1.6in]{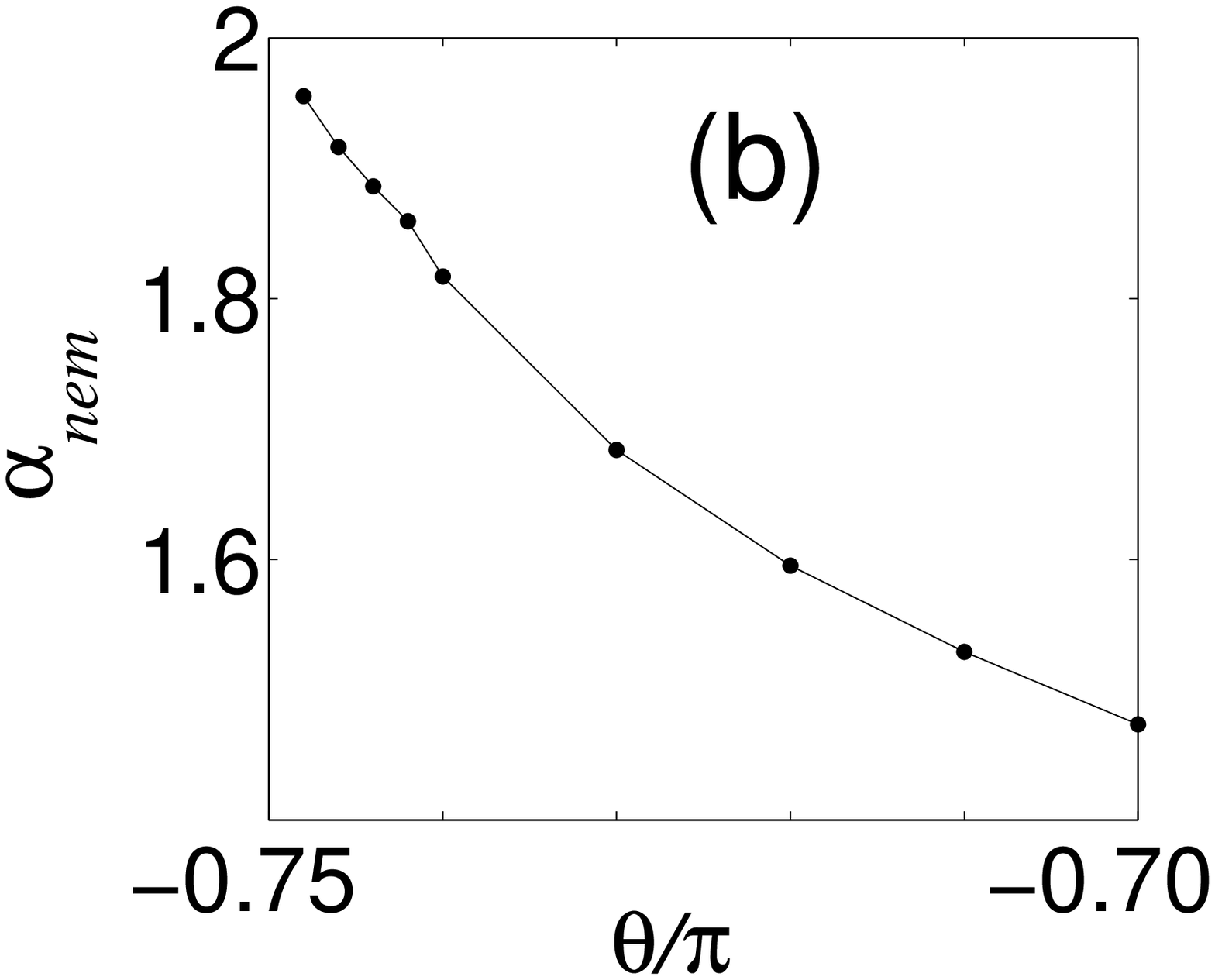}
  \caption{(a) Extrapolated dimer order parameter.
 (b) Exponent of the squared quantum nematic order parameter.  In both cases $8 < N < 36$, and $D=12$ ($D=14$ for $N > 28$).}
\label{fig3}
\end{figure}

\begin{figure}[t]
  \center
  \includegraphics[width=1.55in]{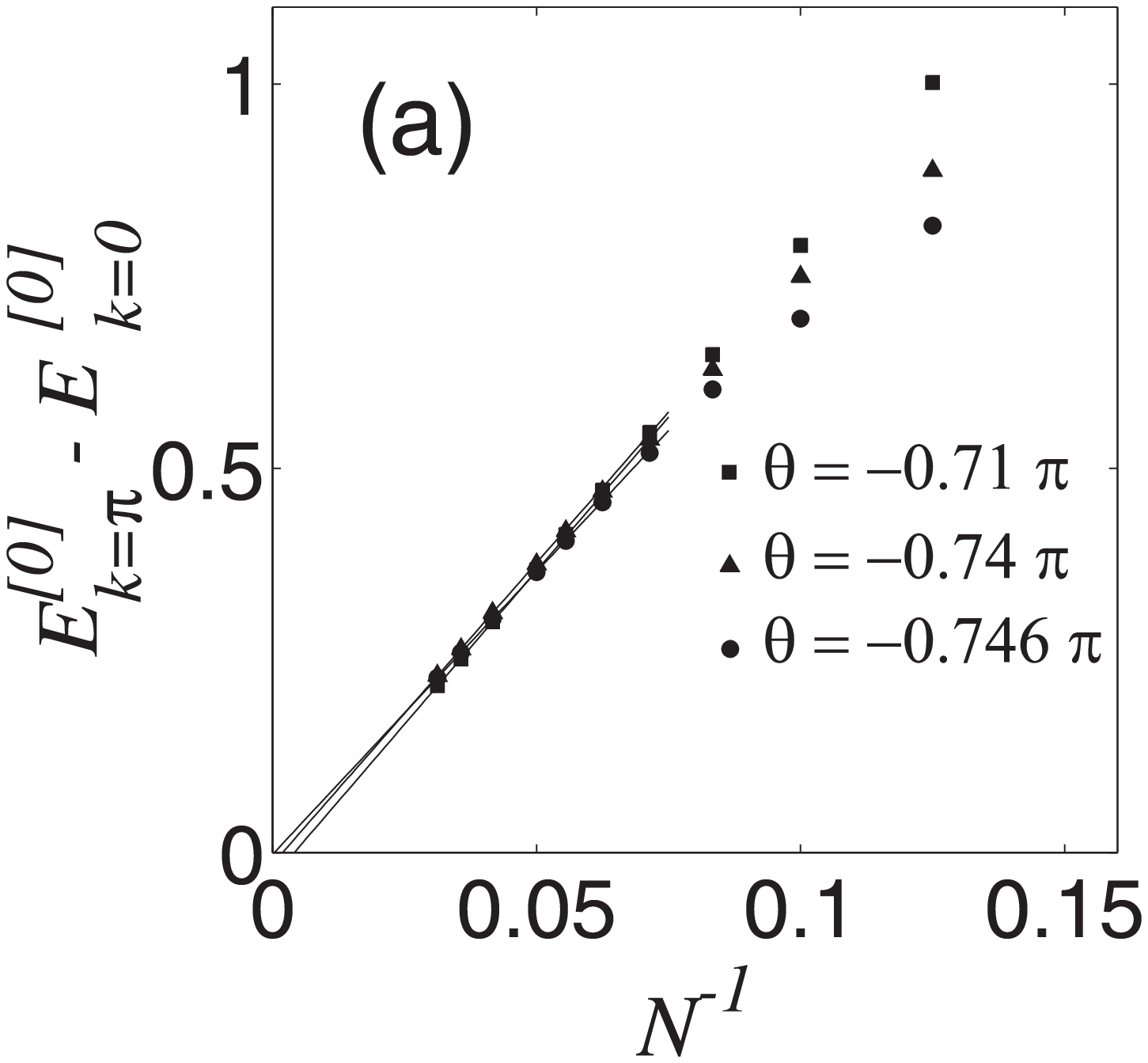}
  \hspace{0.1cm}
  \includegraphics[width=1.5in]{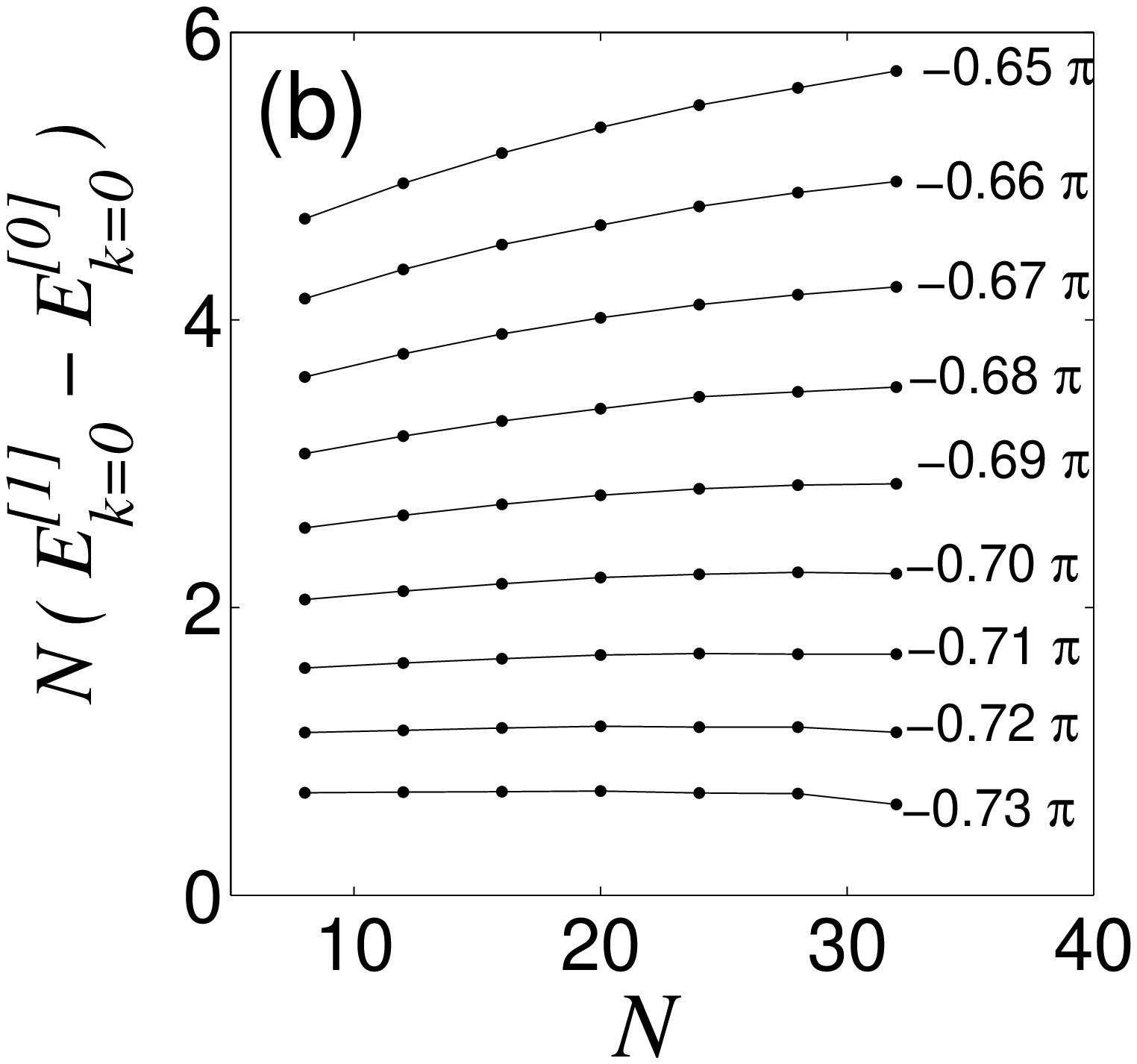}
  \caption{(a) Finite size scaling of the gap between the lowest $k=0$, and $k=\pi$ states. The lines are the extrapolation to $N \rightarrow \infty$. (b) Scaled gap between the lowest energy $k=0$ states. In both plots $D=12$ ($D=14$, for $N=32$), and the error in $E_k$ is $= 10^{-3}$.}
  \label{fig.4}
\end{figure}

\onecolumngrid

\newpage

\begin{center}
{ \large \bf Appendix: Initialization and actualization of operators.}
\end{center}

{\it 
In this appendix we provide a few details for the efficient implementation of our numerical algorithm. We show that by following our procedure for initialization and actualization of operators, the time required by the algorithm in chains with $N$ sites scales like $N^2$.}

\maketitle
%
%Our method relies on the use of a variational wave--function which can be written as a linear combination of translations of Matrix Product States (MPS) with definite linear momentum \cite{paper}:%

Let us recall the calculation of averages with the variational--wave function defined by Eq. (\ref{MPS1}).
$O_{s'\hspace{-0.1cm}, s} = \langle s' | O | s \rangle$ are the matrix elements of a single site observable. We define the following matrices of size $D^2$:
\be
E_{O}^{n,d} = \sum_{s,s'}
\left( A^{s}_{[n]} \otimes (A^{s'}_{[n-d]})^{*} \right) O_{s'\hspace{-0.1cm}, s} ,
\label{positive.maps}
\ee
with the following definition of tensor product:
\be
\left( A \otimes B \right)_{(\alpha,\alpha'),(\beta,\beta')} = A_{\alpha,\beta} B_{\alpha',\beta'}.
\label{positive.maps}
\ee
The expectation value of any operator is a linear combination of $N$ products of $D^2 \times D^2$ matrices:
\be
\langle O_1 O_2 \ldots O_N \rangle = \frac{1}{N} \sum_{n,d} e^{-i k d}
\textmd{tr} \{ E_{O_1}^{n,d} E_{O_2}^{n+1,d} \ldots E^{n-1,d}_{O_N} \}   
\label{mean.expectation}
\ee
The purpose of this appendix is to show how to calculate efficiently the bilinear forms that correspond to the energy and norm of the variational wave--function as a function of a given $A_{[n]}$. These bilinear forms are determined by the relations:
\be
\langle \Psi_k | \Psi_k \rangle = 
\frac{1}{N} \sum_{d} e^{-i k d}
\textmd{tr} \{ E_{\one}^{1,d} E_{\one}^{2,d} \ldots E^{N,d}_{\one} \}  = 
\sum_{\substack{s,s' \\ \alpha,\alpha',\beta,\beta'}}
({A_{[n]}^{s}})^*_{\alpha,\beta} {\N}^{s,s'}_{\alpha,\beta,\alpha',\beta'}[n] (A^{s'}_{[n]})_{\alpha',\beta'}
\nonumber \\
\langle \Psi_k | H | \Psi_k \rangle = 
\frac{1}{N} \sum_{m,d,\mu} e^{-i k d}
\textmd{tr} \{ E_{\sigma_\mu}^{m,d} E_{\sigma_\mu}^{m+1,d} \ldots E^{m-1,d}_{\one} \}  = \sum_{\substack{s,s' \\ \alpha,\alpha',\beta,\beta'}}
({A_{[n]}^{s}})^*_{\alpha,\beta} {\H}^{s,s'}_{\alpha,\beta,\alpha',\beta'}[n] (A^{s'}_{[n]})_{\alpha',\beta'}
\label{bilinear.norm}
\ee   
According to Eq. (\ref{mean.expectation}) the matrices $\H[n]$ and $\N[n]$ in Eq. (\ref{bilinear.norm}) can be expressed as a linear combination of Fourier components:
\be
{\N}[n] = \sum_d e^{- i k d} {\N}[n,d], \ \ {\H}[n] = \sum_d e^{- i k d} {\H}[n,d], 
\label{norm.k}
\ee
where $\N[n,d]$, $\H[n,d]$ depend on matrices $E_O^{m,d}$ only. Remember that at step $n$ we have to solve the generalized eigenvalue problem defined by $\H[n]$, $\N[n]$, to get the optimum $A^s_{[n]}$, and then move to $A^s_{[n+1]}$, and calculate $\N[n]$, $\H[n]$. We repeat this process from $n = 1$ to  $n = N$, that is, we perform a sweep. Our numerical results shows that a number of 10 sweeps is usually enough to converge to the minimum. 

We define the set of products: 
\be
e^{n,m}_d &=& E^{n,d}_\one E^{n+1,d}_\one \ldots E^{m,d}_\one , \nonumber \\
s_{\mu,d}^{n,m} &=& E^{n,d}_{\sigma^\mu} E^{n+1,d}_\one \ldots E^{m,d}_\one , \nonumber \\
t_{\mu,d}^{n,m} &=& E^{n,d}_\one E^{n+1,d}_\one \ldots E^{m,d}_{\sigma^{\mu}} , \nonumber \\
h_d^{n,m} &=& \sum_\mu g_\mu 
\left( E^{n,d}_{\sigma^\mu} E^{n+1,d}_{\sigma^\mu} \ldots E^{m,d}_\one + \ldots
 +  E^{n,d}_\one \ldots E^{m-1,d}_{\sigma^\mu} E^{m,d}_{\sigma^\mu} \right) .
\label{products}
\ee
In terms of (\ref{products}) we find the following expression for the norm:
\be
\N[n,d]^{s,s'}_{\alpha,\beta,\alpha',\beta'}  =
\tr \{
\left( \chi[^{\alpha'}_{\beta'}] \otimes (A^{s'}_{[n-d]})^* \! \right) e^{n+1,n+d-1}_d
\left( A^{s}_{[n+d]} \otimes \chi[^{\alpha}_{\beta}] \right) e^{n+d+1,n-1}_d \},
\label{bilinear.norm.k}
\ee
and the effective Hamiltonian:
\be
& & \hspace{-.5cm} \H[n,d]^{s,s'}_{\alpha,\beta,\alpha',\beta'} = 
\nonumber \\
& &
\tr \{ 
\left( \chi[^{\alpha'}_{\beta'}] \otimes (A^{s'}_{[n-d]})^* \! \right)  \ 
h^{n+1,n+d-1}_d  
\left( A^{s}_{[n+d]} \otimes \chi[^{\alpha}_{\beta}] \right) 
e^{n+d+1,n-1}_d \} +
\nonumber \\
& &
\tr \{
\left( \chi[^{\alpha'}_{\beta'}] \otimes (A^{s'}_{[n-d]})^* \! \right)  \ 
e^{n+1,n+d-1}_d  
\left( A^{s}_{[n+d]} \otimes \chi[^{\alpha}_{\beta}] \right) 
h^{n+d+1,n-1}_d \} +
\nonumber \\
& &
\sum_\mu g_\mu \tr \{
\left( \chi[^{\alpha'}_{\beta'}] \otimes \sum_{t} (A^{t}_{[n-d]})^* \sigma^\mu_{t,s'} \! \right)  \ 
s^{n+1,n+d-1}_{\mu,d}  
\left( A^{s}_{[n+d]} \otimes \chi[^{\alpha}_{\beta}] \right) 
e^{n+d+1,n-1}_d \} +
\nonumber \\
& &
\sum_\mu g_\mu \tr \{
\left( \chi[^{\alpha'}_{\beta'}] \otimes \sum_{t} (A^{t}_{[n-d]})^* \sigma^\mu_{t,s'} \! \right)  \ 
e^{n+1,n+d-1}_d  
\left( A^{s}_{[n+d]} \otimes \chi[^{\alpha}_{\beta}] \right) 
t^{n+d+1,n-1}_{\mu,d} \} +
\nonumber \\
& &
\sum_\mu g_\mu \tr \{
\left( \chi[^{\alpha'}_{\beta'}] \otimes (A^{s'}_{[n-d]})^* \! \right)  \ 
e^{n+1,n+d-1}_{d}  
\left( \sum_t A^{t}_{[n+d]} \sigma^\mu_{t,s} \otimes \chi[^{\alpha}_{\beta}] \right) 
s^{n+d+1,n-1}_{\mu,d} \} +
\nonumber \\
& &
\sum_\mu g_\mu \tr \{
\left( \chi[^{\alpha'}_{\beta'}] \otimes (A^{s'}_{[n-d]})^* \! \right)  \ 
t^{n+1,n+d-1}_{\mu,d}  
\left( \sum_t A^{s}_{[n+d]} \sigma^\mu_{t,s} \otimes \chi[^{\alpha}_{\beta}] \right) 
e^{n+d+1,n-1}_d \}.
\label{bilinear.hamiltonian.k}
\ee
In (\ref{bilinear.norm.k}, \ref{bilinear.hamiltonian.k}) $\chi[^{\alpha}_{\beta}]$ is a $D \times D$ matrix with all elements $0$, but a $1$ in the entry $(\alpha, \beta)$.
Note that we have just substituted $A^s_{[n]}$ by $\chi$, wherever $A^s_{[n]}$ appears in (\ref{bilinear.norm.k}, \ref{bilinear.hamiltonian.k}). These expressions are equivalent to the effective Norm and Hamiltonian operators of DMRG. In principle, $\H[n,d]$, and $\N[n,d]$ could be calculated at each step by means of (\ref{positive.maps}). This would imply to multiply $N^3$ times $E_O$ matrices (corresponding to $N$ sites, $N$ interacting terms in the Hamiltonian, and $N$ values of $d$).
Since the calculation of $\H[n,d]$, and $\N[n,d]$ in terms of the set of matrices $A^s_{[m]}$, $m \neq n$, is the most time demanding part of our algorithm, it is important to find a way to reduce this number of operations. Indeed, the number of multiplications can be reduced to scale like $N$ by storing and actualizing the matrices (\ref{products}) in a way that resembles the procedure for building block operators in DMRG. Let us see how this procedure works.

\begin{figure}[t]
\center
\includegraphics[width=4.in]{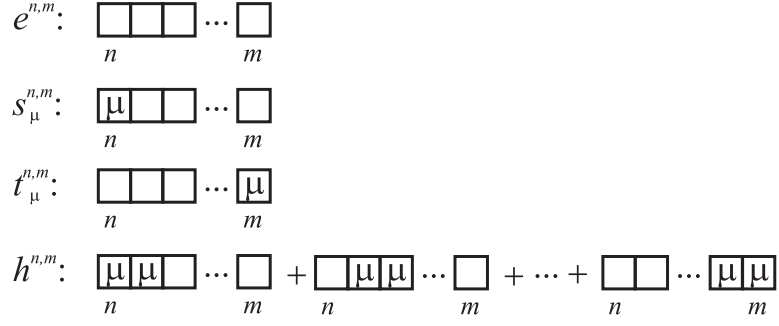}
\caption{Representation of effective operators for the calculation of effective Hamiltonian and Norm operators (index $d$ is omitted).}
\label{Addendum1.fig}
\end{figure}

\begin{figure}[t]
\center
\includegraphics[width=2.5in]{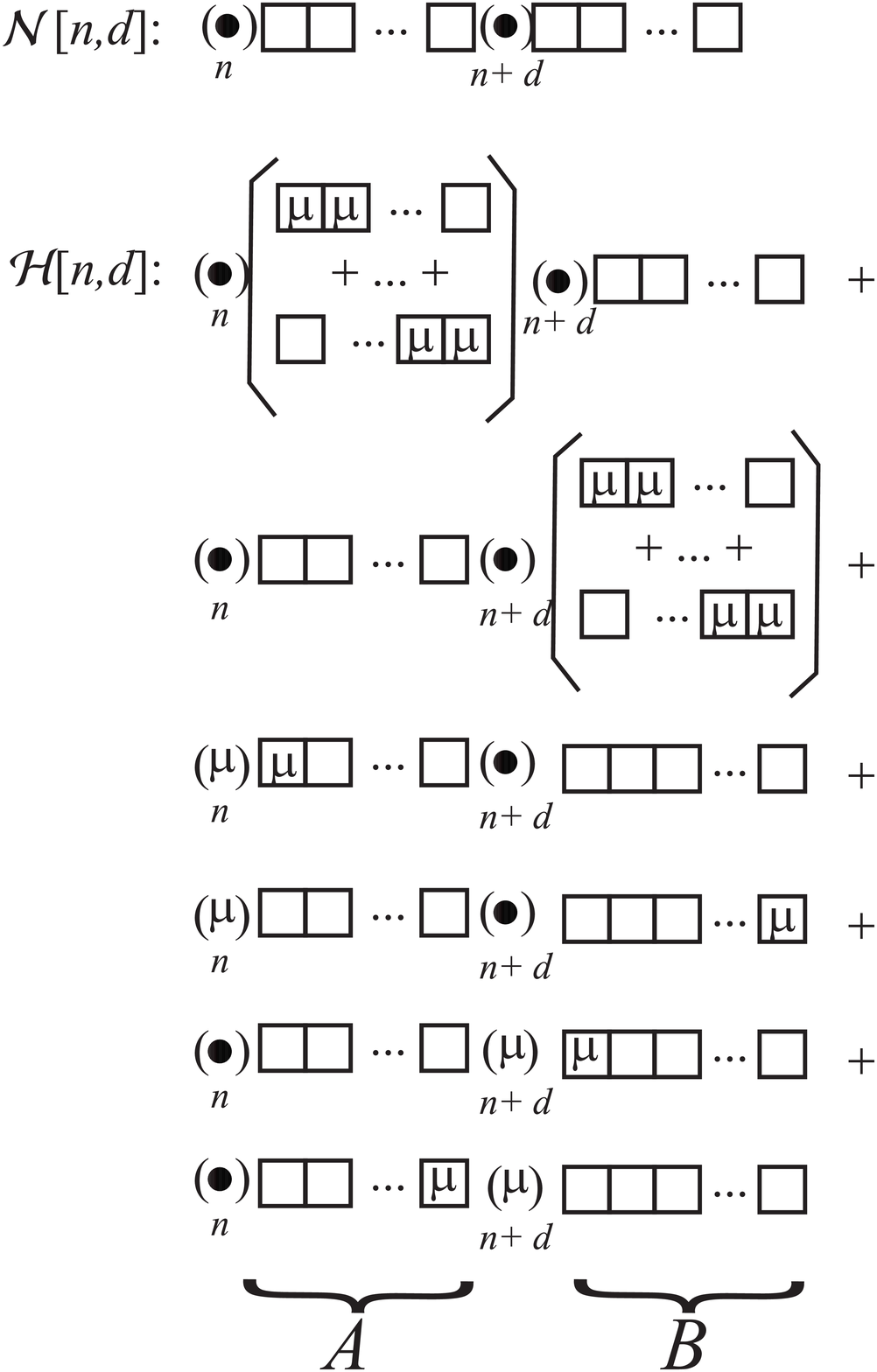}
\caption{Representation of Eqs. (\ref{bilinear.norm.k},\ref{bilinear.hamiltonian.k}). Parenthesis represent sites where matrices $A^s_{[n]}$ appear in Eq. (\ref{bilinear.norm}). Black circles corresponds to terms of the form $\chi \otimes (A^s_{[n]})^*$, $A^s_{[n]} \otimes \chi$, whereas $(\mu)$ corresponds to contractions with the operator $\sigma^\mu$: $\sum_s  \chi \otimes (A^t_{[n]})^* \sigma^\mu_{t,s}$, $\sum_s  (A^t_{[n]}) \sigma^\mu_{t,s} \otimes \chi$.}
\label{Addendum2.fig}
\end{figure}

First of all, we introduce a pictorial representation of Eqs. (\ref{bilinear.norm.k}, \ref{bilinear.hamiltonian.k}). In Fig. \ref{Addendum1.fig} we represent the operators (\ref{products})
that we need to build $\H[n,d]$ and $\N[n,d]$, at each step in the minimization of the energy, see Fig.  \ref{Addendum2.fig}. 
Our task is to calculate efficiently these sets of products by using results of the previous steps, and recursive relations between operators. We notice that one needs two sets of products starting at sites $n$ and $n+d$, which corresponds to each of the left and right terms in Fig. \ref{Addendum2.fig}. In the following we will borrow the DMRG terminology and call these two parts blocks $A$ and $B$. The procedure that we explain below has to be carried out independently for each of the blocks $A$, $B$.

\begin{figure}[t]
\center
\includegraphics[width=4.in]{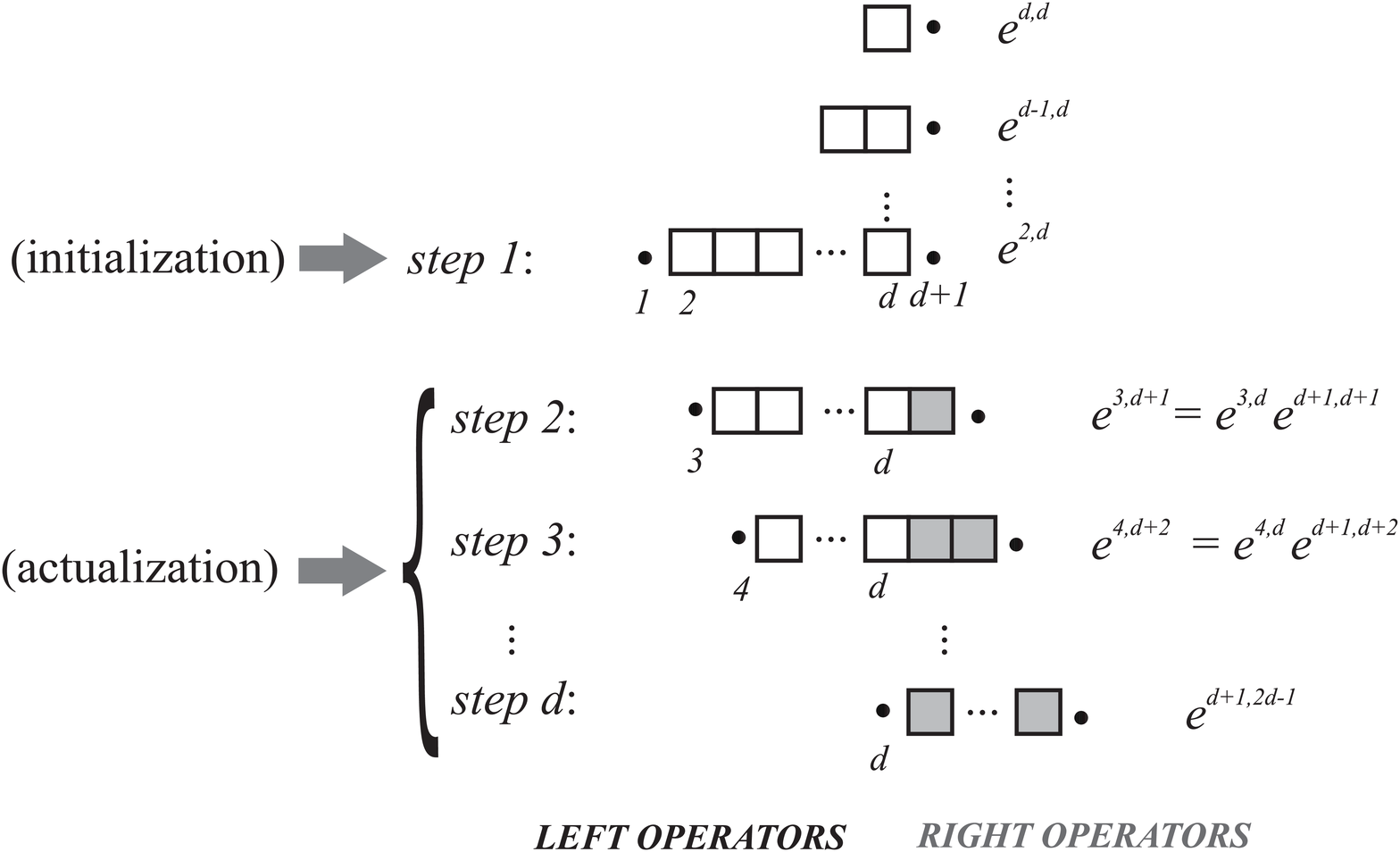}
\caption{Representation of the procedure for initialization/actualization of operators in the calculation of $e^{n+1,n+d-1}$}
\label{Addendum3.fig}
\end{figure}

\begin{figure}[t]
\center
\includegraphics[width=3.in]{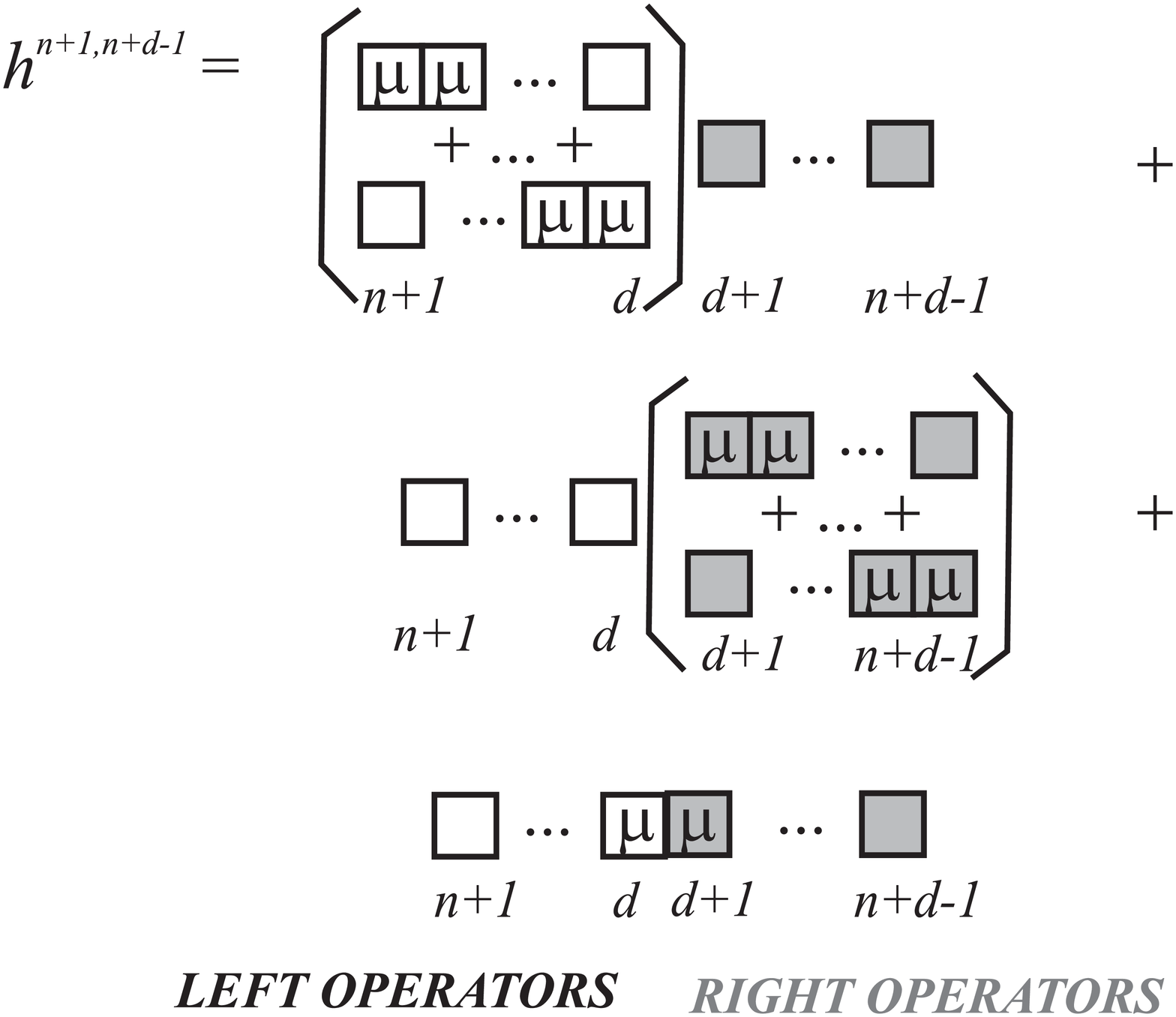}
\caption{Calculation of the operator $h^{n+1,n+d-1}$ by using 'left' and 'right' operators.}
\label{Addendum4.fig}
\end{figure}
Let us consider, for example, the case of $e^{n+1,n+d-1}_d$ appearing in the $A$ block of (\ref{bilinear.hamiltonian.k}).
In the first step, that is, $n=1$, we calculate and store a set of 'left operators' $e^{n',d}_d$, with $n'$ = $2, \ldots, d$, something that can be done in $d-1$ steps.
After solving the eigenvalue problem for $A^s_{[1]}$, we move to the following sites, $n=2,\ldots,d$. As we calculate the new matrices, we actualize (but do not need to store), 'right operators' at each step $n$, $e^{d+1,d-n-1}_d$, by using the 'right operators' of the previous step $n-1$. At each step 'left' and 'right operators' are combined to calculate the matrices that we need for (\ref{bilinear.hamiltonian.k}): $e^{n+1,n+d-1}_d = e^{n+1,d}_d e^{d+1,n+d-1}$ (see Fig. \ref{Addendum3.fig}). The same procedure and decomposition in 'left' and 'right' operators has to be carried out for the products $e^{n+d+1,n-1}_d$, which appear in block $B$ in Eqs. (\ref{bilinear.norm.k}, \ref{bilinear.hamiltonian.k}) (see also Fig. \ref{Addendum2.fig}).

Note that operators carry here and additional index $d$, corresponding to each of the Fourier components in (\ref{norm.k}). The number of operations required for actualizing $\N[n,d]$ at a given step is independent on $N$. On the other hand we have to calculate $N$ contributions $\N[n,d]$ to get $\N[n]$ by means of Eq. (\ref{norm.k}), so that the number of operations per step scales finally like $N$. On the other hand, each sweep takes $N$ steps, so that the whole algorithm scales like $N^2$.

A procedure with the same scaling with $N$ can be applied in the calculation of the Hamiltonian at each step $n$. We explain here, as an example, how to build the operators $h^{n+1,n+d-1}_d$, which appears in block $A$ of the expression (\ref{bilinear.hamiltonian.k}).
In the first step, $n=1$, we calculate and store all the 'left blocks' $h^{n',d}_d$, $e^{n',d}_d$, $s^{n',d,}_{\mu,d}$, with $n'$ = $2, \ldots, d$.  
For this calculation we need to perform a number of multiplications proportional to $d$, since we can use the following recursion relations: $e^{n',d}_d = E^{n',d}_\one e^{n'-1,d}_d$, 
$s^{n',d}_d = E^{n',d}_{\sigma^\mu} e^{n'-1,d}_d$, 
and $h^{n',d}_d = \sum_\mu g_\mu E^{n',d}_{\sigma^\mu} s^{n'-1,d}_{\mu,d} + E^{n',d}_\one h^{n'-1,d}_d$.
After solving the eigenvalue problem for $A^s_{[1]}$, we move to the following sites, $n=2,\ldots,1-d$. 
In the following steps we actualize (but do not need to store), 'left blocks' at each step $n$, $e^{d+1,n+d-1}_d$, $t^{d+1,n+d-1}_{\mu,d}$, $h^{d+1,n+d-1}$, by using the 'left blocks' of the previous step $n-1$. At each step 'left' and 'right operators' are combined to calculate the matrices that we need for (\ref{bilinear.hamiltonian.k}): $h^{n+1,n+d-1}_d = h^{n+1,d}_d \ e^{d+1,n+d-1}_d + e^{n+1,d}_d \ h^{d+1,n+d-1}_d + \sum_\mu g_\mu t^{n+1,d}_{\mu,d} s^{d+1,n+d-1}_{\mu,d}$ (Fig. \ref{Addendum4.fig}).
 
%This procedure for initialization/actualization of operators is carried out independently in the calculation of each Fourier component $\H[n,d]$, $\N[n,d]$, such that the number of matrix multiplications necessary for a step corresponding to finding the optimal value of a matrix $A^s_{[n]}$ scales like $N$, from the $N$ Fourier components. For a sweep along the chain we need $N$ optimizations, thus, the whole algorithm scales like $N^2$.

%


\begin{thebibliography}{99}
\bibitem{note0} For example, $| \Psi_{aux} \rangle$ can be the critical ground state of an exactly solvable model, such that local variables in the $| \Psi_{PEMS} \rangle$ are long--range correlated.
\bibitem{note1}  
$| M \rangle_k$ is equivalent to the multipartite GHZ state,
$(1/\sqrt{N}) \left( |1 \ldots 1 \rangle + |2 \ldots 2 \rangle + \ldots |N \ldots N \rangle \right) $, up to local unitary operations.
\end{thebibliography}
\end{document}